\documentclass[journal]{IEEEtranTIE}

\usepackage{graphicx}
\usepackage{cite}
\usepackage{picinpar}
\usepackage{amsmath}
\usepackage{url}
\usepackage{flushend}
\usepackage[latin1]{inputenc}
\usepackage{colortbl}
\usepackage{soul}
\usepackage{multirow}
\usepackage{pifont}
\usepackage{color}
\usepackage{alltt}
\usepackage[hidelinks]{hyperref}
\usepackage{enumerate}
\usepackage{siunitx}
\usepackage{breakurl}
\usepackage{epstopdf}
\usepackage{pbox}
\usepackage{algorithm}
\usepackage{algorithmic}

\ifCLASSOPTIONcompsoc
  \usepackage[caption=false,font=normalsize,labelfont=sf,textfont=sf]{subfig}
\else
  \usepackage[caption=false,font=footnotesize]{subfig}
\fi

\usepackage{amsmath,amssymb,amsfonts}
\usepackage{cite}
\usepackage{algorithm}
\usepackage{algorithmic}
\usepackage{graphicx}
\usepackage{lipsum}
\usepackage{amsthm}

\begin{document}
\newtheorem{definition}{Definition}
\newtheorem{lemma}{Lemma}
\newtheorem{theorem}{Theorem}
\newtheorem{assumption}{Assumption}
\newtheorem{remark}{Remark}

\title{Recurrent Model Predictive Control}


		

\maketitle
	
\begin{abstract}
This paper proposes an offline control algorithm, called Recurrent Model Predictive Control (RMPC), in order to solve large-scale nonlinear finite-horizon optimal control problems.
As an enhancement of traditional Model Predictive Control (MPC) algorithms, it can adaptively select appropriate model prediction horizon according to current computing resources, so as to improve the policy performance. Our algorithm employs a recurrent function to approximate the optimal policy, which maps the system states and reference values directly to the control inputs. The output of the learned policy network after $N$ recurrent cycles corresponds to the nearly optimal solution of $N$-step MPC.
A policy optimization objective is designed by decomposing the MPC cost function according to the Bellman's principle of optimality. The optimal recurrent policy can be obtained by directly minimizing the designed objective function, which is applicable for general nonlinear and non input-affine systems. 
The hardware-in-the-Loop (HIL) experiment is performed to demonstrate its generality and efficiency. Results show that RMPC is over 5 times faster than the traditional MPC algorithm under identical problem scale.
\end{abstract}


\markboth{IEEE TRANSACTIONS ON INDUSTRIAL ELECTRONICS,~Vol.~xx, No.~xx, September~2020}%
{Shell \MakeLowercase{\textit{et al.}}: Bare Demo of IEEEtran.cls for IEEE Journals}

\definecolor{limegreen}{rgb}{0.2, 0.8, 0.2}
\definecolor{forestgreen}{rgb}{0.13, 0.55, 0.13}
\definecolor{greenhtml}{rgb}{0.0, 0.5, 0.0}

\section{Introduction}
\IEEEPARstart{M}{odel} Predictive Control (MPC) is a well-known method to solve finite-horizon optimal control problems online, which has been extensively investigated in various fields \cite{qin2003survey,vazquez2014model,li2014fast}. 
However, existing MPC algorithms still suffer from a major challenge: relatively low computation efficiency \cite{lee2011model}.

One famous approach to tackle this issue is the moving blocking technique, which assumes constant control input in a fixed portion of the prediction horizon. It increases the computation efficiency by reducing the number of variables to be optimized \cite{cagienard2007move}. However, this solution cannot guarantee the system stability and constraint satisfaction. 
In addition, Wang and Boyd (2009) proposed an early termination interior-point method to reduce the calculation time by limiting the maximum number of iterations per time step \cite{wang2009fast}.

However, these methods are still unable to meet the online computing requirement for nonlinear and large-scale systems. Some control algorithms choose to calculate an near-optimal explicit policy offline, and then implement it online. Bemporad \emph{et al}. (2002) first proposed the explicit MPC method to increase the computation efficiency, which partitioned the constrained state space into several regions and calculated explicit feedback control laws for each region \cite{bemporad2002explicit}. During online implementation, the on-board computer only needs to choose the corresponding state feedback control law according to the current system state, thereby reducing the burden of online calculation to some extent. Such algorithms are only suitable for small-scale systems, since the required storage capacity grows exponentially with the state dimension \cite{kouvaritakis2002needs}. 

Furthermore, significant efforts have been devoted to approximation MPC algorithms, which can reduce polyhedral state regions and simplify explicit control laws. Geyer \emph{et al}. (2008) provided an optimal merging approach to reduce partitions via merging regions with the same control law \cite{geyer2008optimal}. Jones \emph{et al}. (2010) proposed a polytopic approximation method using double description and barycentric functions to estimate the optimal policy, which greatly reduced the partitions and could be applied to any convex problem \cite{jones2010polytopic}. 
Wen \emph{et al}. (2009) proposed a piecewise continuous grid function to represent explicit MPC solution, which reduced the requirements of storage capacity and improve online computation efficiency\cite{wen2009analytical}. Borrelli \emph{et al}. (2010) proposed an explicit MPC algorithm which can be executed partially online and partially offline\cite{borrelli2010computation}. In addition, some MPC studies employed a parameterized function to approximate the MPC controller. They updated the function parameters by minimizing the MPC cost function with a fixed prediction horizon through supervised learning or reinforcement learning \cite{aakesson2005neural,aakesson2006neural,cheng2015neural,duan2019deep}.

Noted that the policy performance and the computation time for each step usually increase with the number of prediction steps. The above-stated algorithms usually have to make a trade-off between control performance and computation time constraints, and select a conservative fixed prediction horizon. While the on-board computation resources are often changing dynamically. These algorithms thus usually lead to calculation timeouts or resources waste. In other words, these algorithms cannot adapt to the dynamic allocation of computing resources and make full use of the available computing time to select the longest model prediction horizon.

In this paper, we propose an offline MPC algorithm, called Recurrent MPC (RMPC), for finite-horizon optimal control problems with large-scale nonlinearities and nonaffine inputs. Our main contributions can be summarized as below:
\begin{enumerate}
	\item A recurrent function is employed to approximate the optimal policy, which maps the system states and reference values directly to the control inputs. Compared to previous algorithms employing non-recurrent functions (such as multi-layer NNs), which must select a fixed prediction horizon previously\cite{aakesson2005neural,aakesson2006neural,cheng2015neural,duan2019deep}, the use of recurrent structure makes the algorithm be able to select appropriate model prediction horizon according to current computing resources. In particular, the output of the learned policy function after $N$ recurrent cycles corresponds to the nearly optimal solution of $N$-step MPC.
	\item A policy optimization objective is designed by decomposing the MPC cost function according to the Bellman's principle of optimality. The optimal recurrent policy can be obtained by directly minimizing the designed objective function. Therefore, unlike the traditional explicit MPC algorithms\cite{bemporad2002explicit, kouvaritakis2002needs,geyer2008optimal,jones2010polytopic,wen2009analytical,borrelli2010computation} that can only handle linear systems, the proposed algorithm is applicable for general nonlinear and non input-affine systems. Meanwhile, the proposed RMPC algorithm utilizes the recursiveness of Bellman's principle. When the cost function of the longest prediction is optimized, the cost function of short prediction will automatically be optimal. Thus the proposed algorithm can deals with different shorter prediction horizons problems while only training with an objective function with respect to a long prediction horizons. Other MPC algorithms \cite{aakesson2005neural,aakesson2006neural,cheng2015neural,duan2019deep,bemporad2002explicit, kouvaritakis2002needs,geyer2008optimal,jones2010polytopic,wen2009analytical,borrelli2010computation}do not consider the recursiveness of Bellman's principle, when the prediction horizons changes, the optimization problem must be reconstructed and the training or computing process must be re-executed to deal with the new problem. 
	\item The proposed RMPC algorithm calculates the optimal control policy previously so only needs to eval the forward inferencing process while using online.Expriments shows that it  is over 5 times faster than the traditional MPC algorithms \cite{Andreas2006Biegler,bonami2008algorithmic} under the same problem scale.
\end{enumerate}

The paper is organized as follows. In Section II, we provide the formulation of the MPC problem. Section III presents RMPC algorithm and proves its convergence. In Section IV, we present simulation demonstrations that show the generalizability and effectiveness of the RMPC algorithm. Section V concludes this paper.

\section{Preliminaries}
Consider general time-invariant discrete-time dynamic system
\begin{equation}
\label{eq.system}
    x_{i+1}=f(x_{i}, u_{i})
\end{equation}
with state $x_{i}\in \mathcal{X} \subset \mathbb{R}^{n}$, control input $u_{i}\in \mathcal{U} \subset \mathbb{R}^{m}$ 
and the system dynamics function $f:\mathbb{R}^{n} \times \mathbb{R}^{m} \to \mathbb{R}^{n}$. We assume that $f(x_i,u_i)$ is Lipschitz continuous on a compact set $\mathcal{X}$, and the system is stabilizable on $\mathcal{X}$.

Define the cost function $V(x_{0},r_{1:N},N)$ of the $N$-step Model Predictive Control (MPC) problem
\begin{equation}
\label{eq.valuedefinition}
    V(x_{0},r_{1:N},N)={\sum_{i=1}^{N}l(x_{i},r_{i},u^{N}_{i-1}(x_{0},r_{1:N}))},
\end{equation}
where $x_{0}$ is initial state, $N$ is length of prediction horizon, $r_{1:N}=[r_{1},r_{2},\cdots,r_{N}]$ is reference trajectory, $V(x_{0},r_{1:N},N)$ is the $N$-step cost function of state $x_{0}$ with reference $r_{1:N}$,  $u^{N}_{i-1}$ is the control input of the $i$th step in $N$-step prediction, and $l\geq0$ is the utility function. The purpose of MPC is to find the optimal control sequence to minimize the objective $V(x_{0},r_{1:N},N)$, which can be denoted as
\begin{equation}
\label{eq.control_sequence}
\begin{aligned}
	\left[{u^{N}_{0}}^*(x_{0},r_{1:N}),{u^{N}_{1}}^*(x_{0},r_{1:N}),\cdots,{u^{N}_{N-1}}^*(x_{0},r_{1:N})\right]
    \\=\mathop{\arg\min}_{u^{N}_{0},u^{N}_{1},\cdots,u^{N}_{N-1}}V(x_{0},r_{1:N},N),
\end{aligned}
\end{equation}
where the superscript $^*$ represents optimal. 
\section{Recurrent Model Predictive Control}
\subsection{Recurrent Policy Function}
In practical applications, we only need to execute the first control input ${u^{N}_{0}}^*(x_{0},r_{1:N})$ of the optimal sequence in \eqref{eq.control_sequence} at each step. Given a control problem, assume that $N_{\text{max}}$ is the maximum feasible prediction horizon. Our aim is to make full use of computation resources and adaptively select the longest prediction horizon $k\in[1,N_{\text{max}}]$, which means that we need to calculate and store the optimal control input ${u^{k}_0}^*(x,r_{1:k})$ of $\forall x\in \mathcal{X}$, $\forall r_{1:k}$ and $\forall k\in[1,N_{\text{max}}]$ in advance. This requires us to find an efficient way to represent the policy and solve it offline.

\begin{figure}[htb]
\captionsetup{justification =raggedright,
              singlelinecheck = false,labelsep=period, font=small}

\centering{\includegraphics[width=0.42\textwidth]{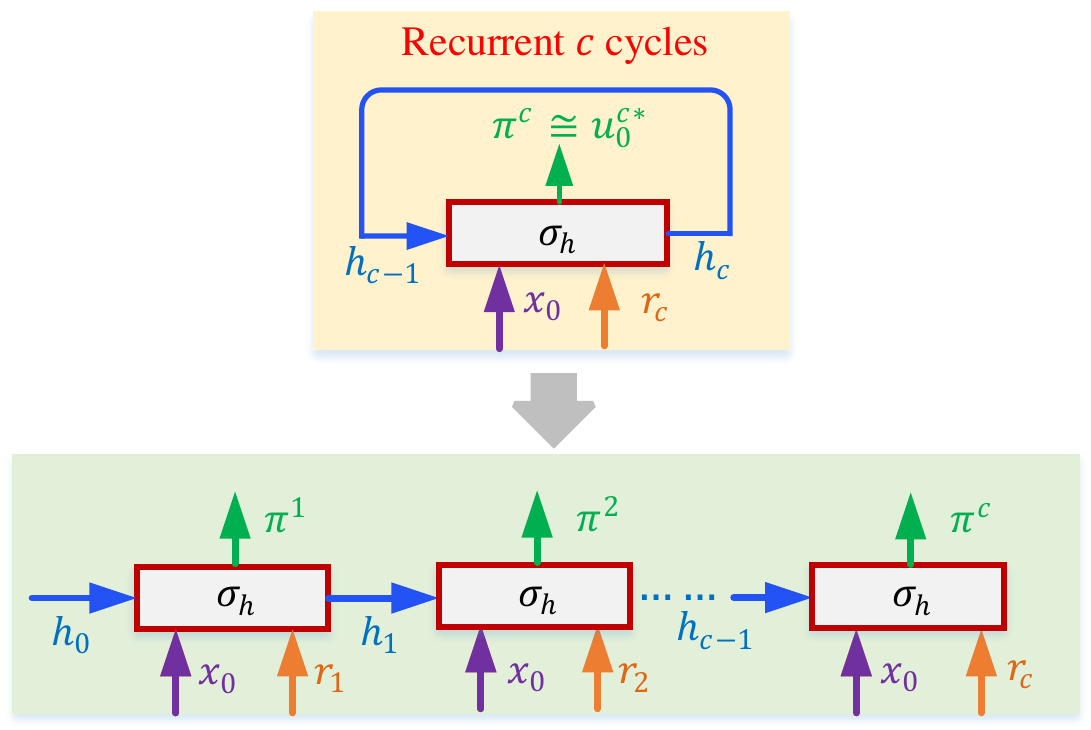}}
\caption{The structure of the recurrent policy function.}
\label{fig_structure}
\end{figure}

We firstly introduce a recurrent function, denoted as $\pi^{c} (x_{0},r_{1:c};\theta)$, to approximate the control input ${u^{c}_{0}}^*(x_{0},r_{1:c})$, where $\theta$ is the vector of function parameters and $c$ is the number of recurrent cycles of the policy function. The goal of the proposed Recurrent MPC (RMPC) algorithm is to find the optimal parameters $\theta^*$, such that
\begin{equation}
\label{eq.equation_optimal}
\begin{aligned}
    \pi^{c}(x_{0},r_{1:c};\theta^*)&={u^{c}_{0}}^*(x_{0},r_{1:c}),\\
    \forall x_{0},r_{1:c} &\in \mathcal{X}, \forall c \in [1,N_{\text{max}}].
\end{aligned}
\end{equation}
The structure of the recurrent policy function is illustrated in Fig. \ref{fig_structure}. All recurrent cycles share the same parameters $\theta$, where $h_c\in\mathbb{R}^q$ is the vector of hidden states.

Each recurrent cycle is mathematically described as 
\begin{equation}
\label{eq.recurrentstructure}
\begin{aligned}
    &h_c=\sigma_h(x_{0},r_{c},h_{c-1};\theta_{h}),
    \\&\pi^c (x_{0},r_{1:c};\theta)=\sigma_y(h_c;\theta_{y}),
    \\&c\in[1,N_{\text{max}}], \theta=\theta_{h}\mathop{\cup}\theta_{y},
\end{aligned}
\end{equation}
where $h_0=0$, $\sigma_h$ and $\sigma_y$ are activation functions of hidden layer and output layer, respectively. 

\begin{figure}[!htb]
\captionsetup{justification =raggedright,
              singlelinecheck = false,labelsep=period, font=small}
\centering{\includegraphics[width=0.5\textwidth]{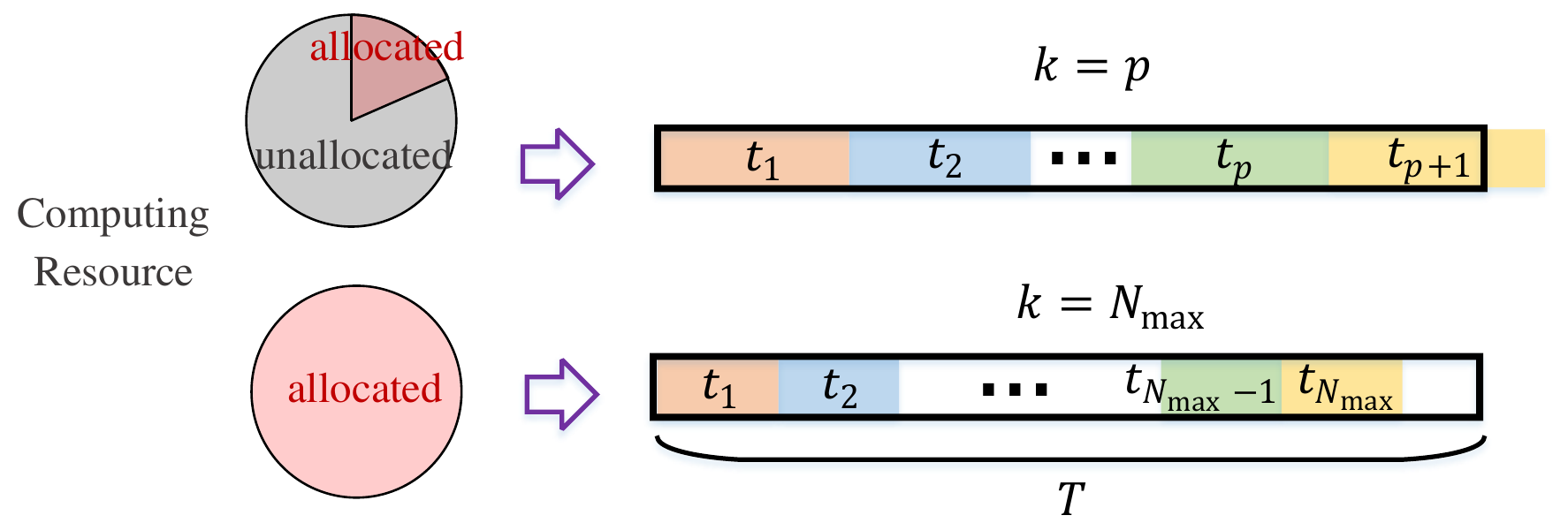}}
\caption{Maximum recurrent cycles in different cases.}
\label{fig_resource}
\end{figure}

As shown in Fig. \ref{fig_structure}, the recurrent policy function calculates and outputs a control input at each recurrent cycle. Assuming that we have found the optimal parameters $\theta^*$, it follows that the output of the $c$th cycle  $\pi^c(x_{0},r_{1:c};\theta^*)={u^{c}_{0}}^*(x_{0},r_{1:c})$ for $\forall c\in[1,N_{\text{max}}]$. This indicates that the more cycles, the longer the prediction horizon. In practical applications, the calculation time of each cycle $t_c$ is different due to the dynamic change of computing resource allocation (see Fig. \ref{fig_resource}). At each time step, the total time assigned to the control input calculation is assumed to be $T$. Denoting the total number of the recurrent cycles at each time step as $k$, then the control input is $\pi^k(x_{0},r_{1:k};\theta^*)$, where
\begin{equation}
\nonumber
k=\left\{
\begin{aligned}
&N_\text{max}, &\quad \sum_{c=1}^{N_{\text{max}}}t_c\le T \\
&p, &\quad \sum_{c=1}^{p}t_c\le T < \sum_{c=1}^{p+1}t_c.
\end{aligned}
\right.
\end{equation}
Therefore, the recurrent policy is able to make full use of computing resources and adaptively select the longest prediction step $k$. In other word, the more computing resources allocated, the longer prediction horizon will be selected, which usually would lead to the better control performance.
\begin{remark}
Previous MPC algorithms employs non-recurrent form neural networks\cite{aakesson2005neural,aakesson2006neural,cheng2015neural,duan2019deep}, which must select a fix prediction horizon previously. RMPC employes recurrent function to approximate the optimal policy, which maps the system states and reference values directly to the control inputs.  The use of recurrent structure makes the algorithm be able to select appropriate model prediction horizon according to current computing resources. The output of the learned policy network after $N$ recurrent cycles corresponds to the nearly optimal solution of $N$-step MPC.
\end{remark}

\subsection{Objective Function}

To find the optimal parameters $\theta^*$ offline, we first need to represent the MPC cost function in \eqref{eq.valuedefinition} in terms of $\theta$, denoted by $V(x_{0},r_{1:N},N; \theta)$. From \eqref{eq.valuedefinition} and the Bellman's principle of optimality, the global minimum $V^*(x_{0},r_{1:N},N)$ can be expressed as:
\begin{equation}
\nonumber
\begin{aligned}
   V^*(&x_{0},r_{1:N},N)
   \\&=l(x_{1},r_{1},{u^{N}_0}^*(x_0,r_{1:N}))+V^*(x_{1},r_{2:N},N-1)\\
   &=\sum_{i=1}^{2}l(x_i,r_i,{u^{N-i+1}_{0}}^*(x_{i-1},r_{i:N})) +\\
   &\qquad\qquad\qquad\qquad \qquad\qquad\qquad V^*(x_{2},r_{3:N},N-2)\\
   &\vdots\\
   &=\sum_{i=1}^{N-1}l(x_i,r_i,{u^{N-i+1}_{0}}^*(x_{i-1},r_{i:N})) +V^*(x_{N-1},r_N, 1)\\
   &=\sum_{i=1}^{N}l(x_i,r_i,{u^{N-i+1}_{0}}^*(x_{i-1},r_{i:N})).
\end{aligned}
\end{equation}
It follows that
\begin{equation}
\label{eq.optimal_V}
\begin{aligned}
V^*(x_{0},r_{1:N},N)&=\sum_{i=1}^{N}l(x_{i},r_{i},{u^{N}_{i-1}}^*(x_{0},r_{1:N}))\\
&=\sum_{i=1}^{N}l(x_i,r_i,{u^{N-i+1}_{0}}^*(x_{i-1},r_{i:N})).
\end{aligned}
\end{equation}
Therefore, for the same $x_0$ and $r_{1:N}$, it is clear that 
\begin{equation}
\label{eq.optimalforanyone}
{u^{N}_{i-1}}^*(x_{0},r_{1:N})={u^{N-i+1}_{0}}^*(x_{i-1},r_{i:N}),\quad \forall i\in[1,N].
\end{equation}
This indicates that the $i$th optimal control input ${u^{N}_{i-1}}^*(x_{0},r_{1:N})$ in \eqref{eq.control_sequence} can be regarded as the optimal control input of the $N$-$i$+$1$-step MPC control problem with initial state $x_{i-1}$. Hence,by replacing all $u^{N}_{i-1}(x_{0},r_{1:N}))$ in \eqref{eq.valuedefinition} with $u^{N-i+1}_{0}(x_{i-1},r_{i:N})$, the $N$-step MPC control problem can also be solved via minimizing $V(x_{0},r_{1:N},N)=\sum_{i=1}^{N}l(x_i,r_i,u^{N-i+1}_{0}(x_{i-1},r_{i:N}))$.  Then, we can obtain the $N$-step cost function in terms of $\theta$:
\begin{equation}
\label{eq.appro_V}
V(x_{0},r_{1:N},N;\theta)=\sum_{i=1}^{N}l(x_i,r_i,\pi^{N-i+1}(x_{i-1},r_{i:N};\theta))
\end{equation}

To find the optimal parameters $\theta^*$ that make \eqref{eq.equation_optimal} hold, we can construct the following objective function:
\begin{equation}
\label{eq.lossfunction}
    J(\theta)=\mathop{\mathbb{E}}_{\substack{x_0\in\mathcal{X}\\r_{1:{N_{\text{max}}}}\in\mathcal{X}}}\Big\{V(x_{0},r_{1:{N_{\text{max}}}},N_{\text{max}};\theta)\Big\}.
\end{equation}
Therefore, we can update $\theta$ by directly minimizing $J(\theta)$. The policy update gradients can be derived as
\begin{equation}   
\label{eq.updatagradient}
\begin{aligned}
\frac{\text{d}J}{\text{d}\theta}&=\mathop{\mathbb{E}}_{
\small\begin{array}{ccc}
\small x_0\in\mathcal{X}\\
\small r_{1:N_{\text{max}}}\in\mathcal{X}\\
\end{array} } \Big\{\frac{\text{d}V(x_{0},r_{1:N_{\text{max}}},N_{\text{max}};\theta)}{\text{d}\theta}\Big\},\\
\end{aligned}
\end{equation}
where
\begin{equation}
\nonumber
\begin{aligned}
&\frac{\text{d}V(x_{0},r_{1:N},N_{\text{max}};\theta)}{\text{d}\theta}=\\
&\qquad\qquad\qquad\sum_{i=1}^{N_{\text{max}}}\frac{\text{d}l(x_{i},r_{i},\pi^{N_{\text{max}}-i+1}(x_{i-1},r_{i:N_{\text{max}}};\theta))}{\text{d} \theta}.
\end{aligned}
\end{equation}
Denoting $\pi^{N_{\text{max}}-i+1}(x_{i-1},r_{i:N_{\text{max}}};\theta)$ as $\pi^{N_{\text{max}}-i+1}$ and $l(x_{i},r_{i},\pi^{N_{\text{max}}-i+1}(x_{i-1},r_{i:N_{\text{max}}};\theta))$ as $l_{i}$, we have
\begin{equation}
\nonumber
\begin{aligned}
&\frac{\text{d} V(x_{0},r_{1:N_{\text{max}}},N_{\text{max}};\theta)}{\text{d} \theta}=\\
&\qquad\qquad\qquad\sum_{i=1}^{N_{\text{max}}}
\Big\{\frac{\partial l_{i}}{\partial x_{i}}\frac{\mathrm{d}x_i}{\mathrm{d}\theta}+\frac{\partial l_{i}}{\partial \pi^{N_{\text{max}}-i+1}}\frac{\mathrm{d}\pi^{N_{\text{max}}-i+1}}{\mathrm{d}\theta}\Big\},
\end{aligned}
\end{equation}

\begin{figure}[!htb]
\captionsetup{justification =raggedright,
              singlelinecheck = false,labelsep=period, font=small}
\centering{\includegraphics[width=0.5\textwidth]{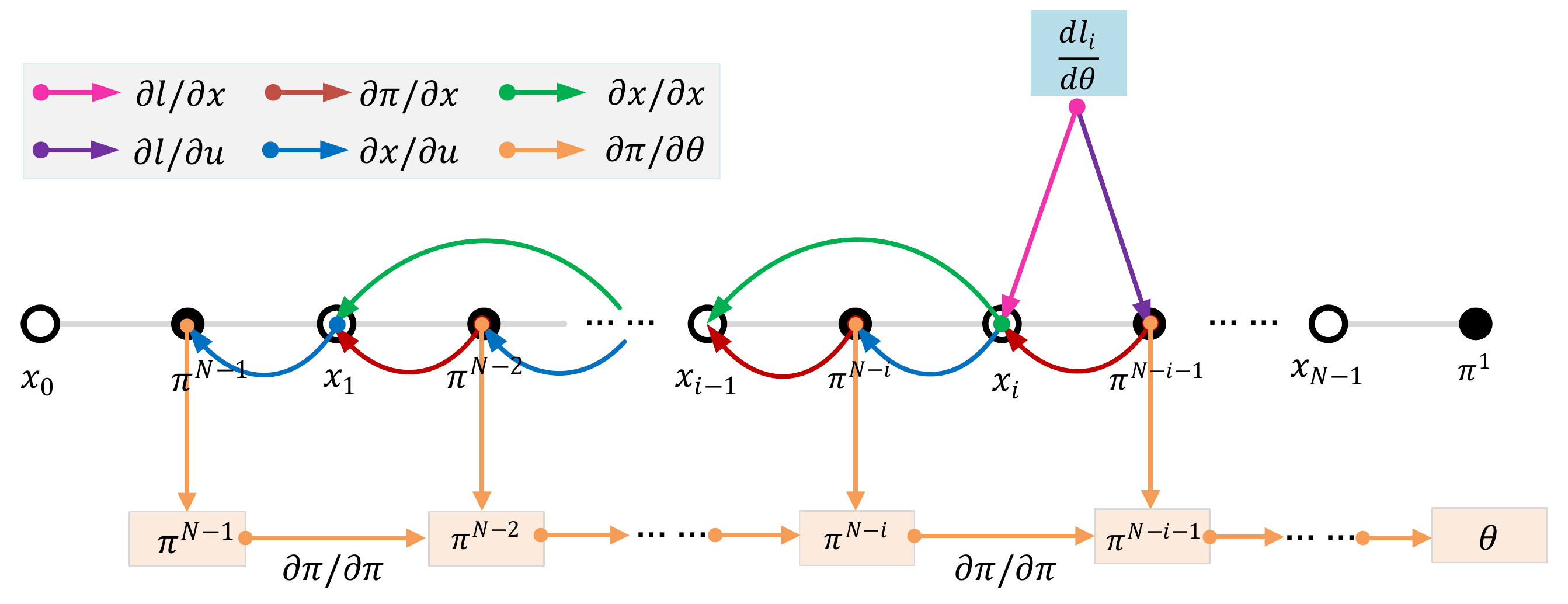}}
\caption{Gradient backpropagation for recurrent function}
\label{fig_gradient}
\end{figure}

By defining two immediate variables, $\phi_{i}=\frac{\mathrm{d}x_i}{\mathrm{d}\theta}$ and $\psi_{i}=\frac{\mathrm{d}\pi^{N_{\text{max}}-i+1}}{\mathrm{d}\theta}$, we have their recursive formula and the details of gradient backpropagation are shown in Fig.\ref{fig_gradient}.

\begin{equation}
\nonumber
\begin{aligned}
\phi_{i} &=\left\{
\begin{aligned}
&0, \qquad \qquad \qquad \qquad \qquad \qquad \qquad \qquad\qquad i=0 \\
&\frac{\partial f(x_{i-1},\pi^{N_{\text{max}}-i+1})}{\partial x_{i-1}}\phi_{i-1} +\\
&\qquad\qquad\qquad\qquad\frac{\partial f(x_{i-1},\pi^{N_{\text{max}}-i+1})}{\partial \pi^{N_{\text{max}}-i+1}}\psi_{i},\quad\text{else}
\end{aligned}
\right. \\
\psi_{i} &=\frac{\partial\pi^{N_{\text{max}}-i+1}}{\partial x_{i-1}}\phi_{i-1} + \frac{\partial\pi^{N_{\text{max}}-i+1}}{\partial \theta}.
\end{aligned}
\end{equation} 

Therefore, the gradient formula can be simplified as:
\begin{equation}
\nonumber
\frac{\text{d} V(x_{0},r_{1:N_{\text{max}}},N_{\text{max}};\theta)}{\text{d} \theta}=\sum_{i=1}^{N_{\text{max}}}
\Big\{\frac{\partial l_{i}}{\partial x_{i}}\phi_i+\frac{\partial l_{i}}{\partial \pi^{N_{\text{max}}-i+1}}\psi_i\Big\},
\end{equation}

\begin{figure}[!htb]
\captionsetup{justification =raggedright,
              singlelinecheck = false,labelsep=period, font=small}
\centering{\includegraphics[width=0.5\textwidth]{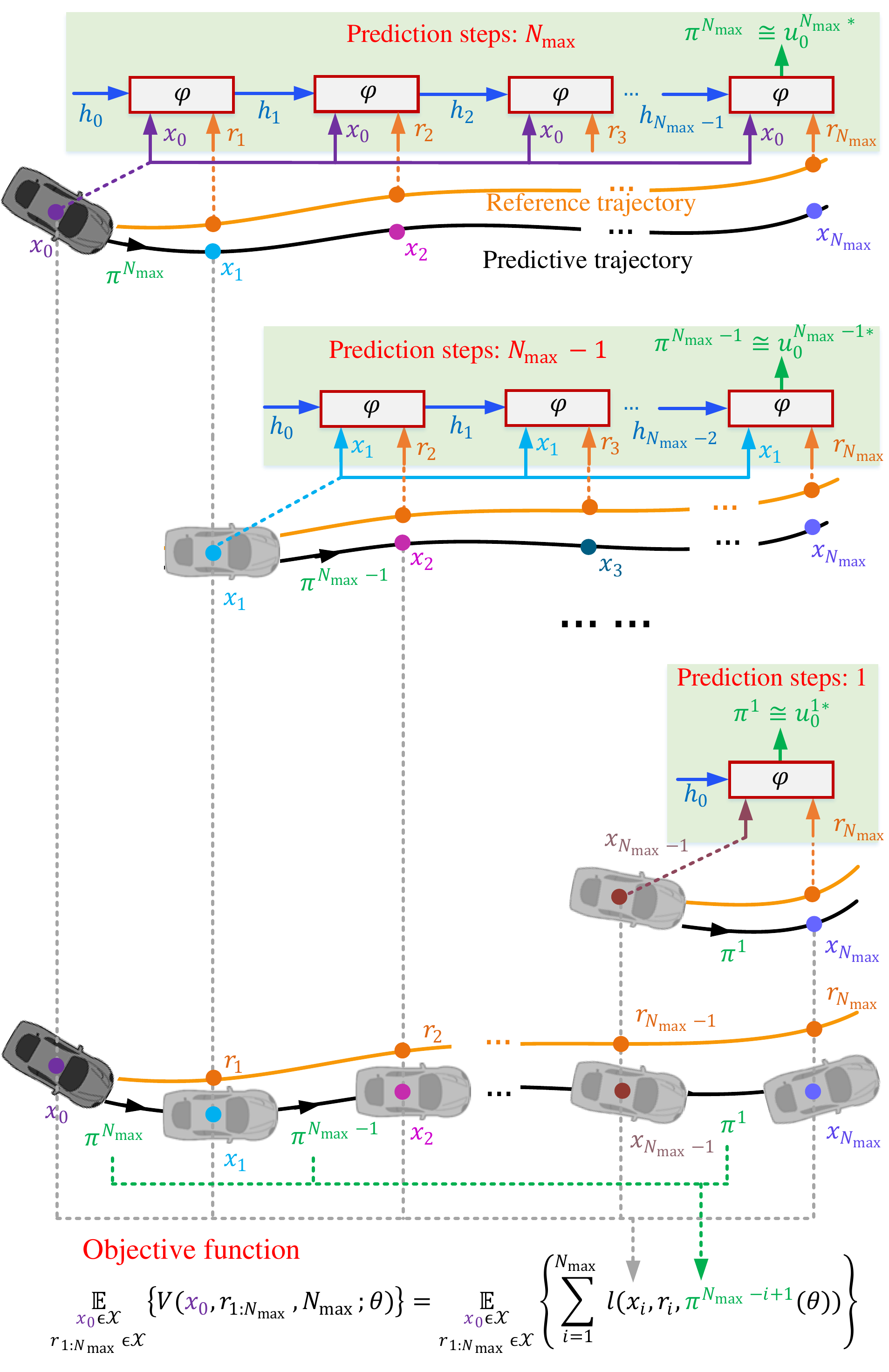}}
\caption{The training flowchart of RMPC algorithm}
\label{fig_obj}
\end{figure}

Taking the Gradient Descent (GD) method as an example, the updating rules of the policy function are
\begin{equation}
\label{eq.update_rule}
\begin{aligned}
\theta_{K+1} &= -\alpha_{\theta} \frac{\text{d}J}{\text{d}\theta} + \theta_K,
\end{aligned}
\end{equation}
where $\alpha_{\theta} $ denotes the learning rate and $K$ indicates $K$th iteration. 

The framework and pseudo-code of the proposed RMPC algorithm is shown in  Algorithm \ref{alg:RMPC} and Fig. \ref{fig_obj}. 

\begin{algorithm}[!htb]
\caption{RMPC algorithm}
\label{alg:RMPC}
\begin{algorithmic}
   \STATE Given an appropriate learning rate $\alpha_\theta$ and any arbitrarily small positive number $\epsilon$.
   \STATE Initial with arbitrary $\theta_0$
\REPEAT 
\STATE Randomly select $x_0, r_{1:N_{\text{max}}}\in \mathcal{X}$
\STATE Calculate $\frac{\text{d}J(\theta_K)}{\text{d}\theta_K}$ using \eqref{eq.updatagradient}
\STATE Update policy function using \eqref{eq.update_rule}
\UNTIL $|J(\theta_{K+1})-J(\theta_{K})|\le \epsilon$
\end{algorithmic}
\end{algorithm}
\begin{remark}
Traditional explicit MPC algorithms\cite{bemporad2002explicit, kouvaritakis2002needs,geyer2008optimal,jones2010polytopic,wen2009analytical,borrelli2010computation} can only handle linear systems.The proposed RMPC algorithm uses an optimization objective designed by decomposing the MPC cost function according to the Bellman's principle of optimality. The optimal recurrent policy can be obtained by directly minimizing the designed \textcolor{red}{objective} function without restrictions on the form of systems.  Meanwhile, the proposed algorithm utilizes the recursiveness of Bellman's principle. When the cost function of the longest prediction is optimized, the cost function of short prediction will automatically be optimal. Thus the proposed algorithm can deals with different shorter prediction horizons problems while only training with an objective function with respect to a long prediction horizons. Other MPC algorithms \cite{aakesson2005neural,aakesson2006neural,cheng2015neural,duan2019deep,bemporad2002explicit, kouvaritakis2002needs,geyer2008optimal,jones2010polytopic,wen2009analytical,borrelli2010computation}do not consider the recursiveness of Bellman's principle, when the prediction horizons changes, the optimization problem must be reconstructed and the training or computing process must be re-executed to deal with the new problem.
\end{remark}
\subsection{Convergence and Optimality}
There are many types of recurrent functions belonging to the structure defined in \eqref{eq.recurrentstructure}, and recurrent neural networks (RNN) are the most commonly used. In recent years, deep RNNs have been successfully implemented in many fields, such as natural language processing and system control, attributing to their ability to process sequential data \cite{mikolov2010recurrent,li2017novel}. Next, we will show that as the iteration index $K\rightarrow\infty$, the optimal policy $\pi^c (x_{0},r_{1:c};\theta^*)$ that make \eqref{eq.equation_optimal} hold can be achieved using Algorithm \ref{alg:RMPC}, as long as $\pi^c (x_{0},r_{1:c};\theta)$ is an over-parameterized RNN. The over-parameterization means that the number of hidden neurons is sufficiently large. Before the main theorem, the following lemma and assumption need to be introduced. 

\begin{lemma} 
\label{lemma.ability}
(Universal Approximation Theorem\cite{li1992approximation,schafer2007recurrent,hammer2000approximation}). Consider a sequence of finite functions $\{F^i(y^i)\}_{i=1}^n$, where $y^i=[y_1,y_2,\hdots,y_i]\in\mathbb{R}^i$, $i$ is the input dimension, $F^i(y^i): \mathbb{R}^i\rightarrow \mathbb{R}^d$ is a continuous function on a compact set and $d$ is the output dimension. Describe the RNN ${G}^c (y^c;W,b)$ as
\begin{equation}
\nonumber
\begin{aligned}
    &h_c=\sigma_h(W_h^\top y^{c}+U_h^\top h_{c-1}+b_h),\\
    &{G}^c (y^c;W,b)=\sigma_y(W_y^\top h_{c}+b_y),
\end{aligned}
\end{equation}
where $c$ is the number of recurrent cycles,  $W=W_h\mathop{\cup}W_y$, $b=b_h\mathop{\cup}b_y$ and $U_h$  are parameters, $\sigma_h$ and  $\sigma_y$ are activation functions.  Supposing ${G}^c (y^c;W,b)$ is over-parameterized, for any $\{F^i(y^i)\}_{i=1}^n$, $\exists U_h, W, b$, such that
\begin{equation}
\nonumber
 \left \| {G}^c(y^c; W,b)-F^c(y^c) \right \|_{\infty} \le \epsilon, \quad \forall c \in [1,n],
\end{equation} 
where $\epsilon \in\mathbb{R^+}$ is an arbitrarily small error.
\end{lemma}

The reported experimental results and theoretical proofs have shown that the straightforward optimization methods, such as GD and Stochastic GD (SGD), can find global minima of most training objectives in polynomial time if the approximate function is an over-parameterized neural network or RNN \cite{allen2019convergence,du2019gradient}. Based on this fact, we make the following assumption.

\begin{assumption} 
\label{assumption.global}
If the approximate function is an over-parameterized RNN, the global minimum of objective function in \eqref{eq.lossfunction} can be found using an appropriate optimization algorithm such as SGD \cite{allen2019convergencernn}.
\end{assumption}

We now present our main result.

\begin{theorem}
\label{theorem.optimality}
(Recurrent Model Predictive Control). Suppose $\pi^c (x_{0},r_{1:c};\theta)$ is an over-parameterized RNN. Through Algorithm \ref{alg:RMPC}, any initial parameters $\theta_0$ will converge to $\theta^*$, such that \eqref{eq.equation_optimal} holds.
\end{theorem}

\noindent \textbf{Proof.} From Assumption \ref{assumption.global}, we can always find $\theta^{\dagger}$  by repeatedly minimizing $J(\theta)$ using \eqref{eq.update_rule}, such that
\begin{equation}
\nonumber
\centering
\theta^\dagger= \arg\min_\theta J(\theta).
\end{equation} 
According to the definition of $J(\theta)$ in \eqref{eq.lossfunction}, we have
\begin{equation}
\nonumber
\begin{aligned}
\min_{\theta}J(\theta)&=\min_{\theta}\mathop{\mathbb{E}}_{\substack{x_0\in\mathcal{X}\\r_{1:{N_{\text{max}}}}\in\mathcal{X}}}\Big\{V(x_{0},r_{1:{N_{\text{max}}}},N_{\text{max}};\theta)\Big\}\\
&\ge\mathop{\mathbb{E}}_{\substack{x_0\in\mathcal{X}\\r_{1:{N_{\text{max}}}}\in\mathcal{X}}}\Big\{\min_{\theta}V(x_{0},r_{1:{N_{\text{max}}}},N_{\text{max}};\theta)\Big\}, \quad \forall \theta.
\end{aligned}
\end{equation}
By Lemma \ref{lemma.ability}, there always $\exists \theta$, such that 
\begin{equation}
\centering
\nonumber
\theta=\arg\min_{\theta}V(x_{0},r_{1:N_{\text{max}}},N_{\text{max}};\theta),\quad \forall x,r_{1:N_{\text{max}}}\in \mathcal{X}.
\end{equation}
Since $\theta^{\dagger}$ is the global minimum of $J(\theta)$, it follows that
\begin{equation}
\centering
\nonumber
J(\theta^{\dagger})=\mathop{\mathbb{E}}_{\substack{x_0\in\mathcal{X}\\r_{1:{N_{\text{max}}}}\in\mathcal{X}}}\Big\{\min_{\theta}V(x_{0},r_{1:{N_{\text{max}}}},N_{\text{max}};\theta)\Big\}, \quad \forall \theta.
\end{equation}
and
\begin{equation}
\centering
\nonumber
\theta^{\dagger}=\arg\min_{\theta}V(x_{0},r_{1:N_{\text{max}}},N_{\text{max}};\theta),\quad\forall x, r_{1:N_{\text{max}}}\in \mathcal{X}.
\end{equation}
Then, according to \eqref{eq.optimal_V}, \eqref{eq.optimalforanyone} and the Bellman's principle of optimality, $\theta^\dagger$ can also make \eqref{eq.equation_optimal} hold, i.e., $\theta^\dagger=\theta^*$. 

Thus, we have proven that RMPC algorithm can converge to $\theta^*$. In other words, it can find the nearly optimal policy of MPC with different prediction horizon, whose output after $c$th recurrent cycles corresponds to the nearly optimal solution of $c$-step MPC.

\section{Algorithm Verification}

In order to evaluate the performance of the proposed RMPC algorithm, we choose the vehicle lateral control problem in path tracking task as an example \cite{li2017driver}. 

\subsection{Overall Settings}
\begin{figure}[!htb]
\captionsetup{
              singlelinecheck = false,labelsep=period, font=small}
\centering{\includegraphics[width=0.5\textwidth]{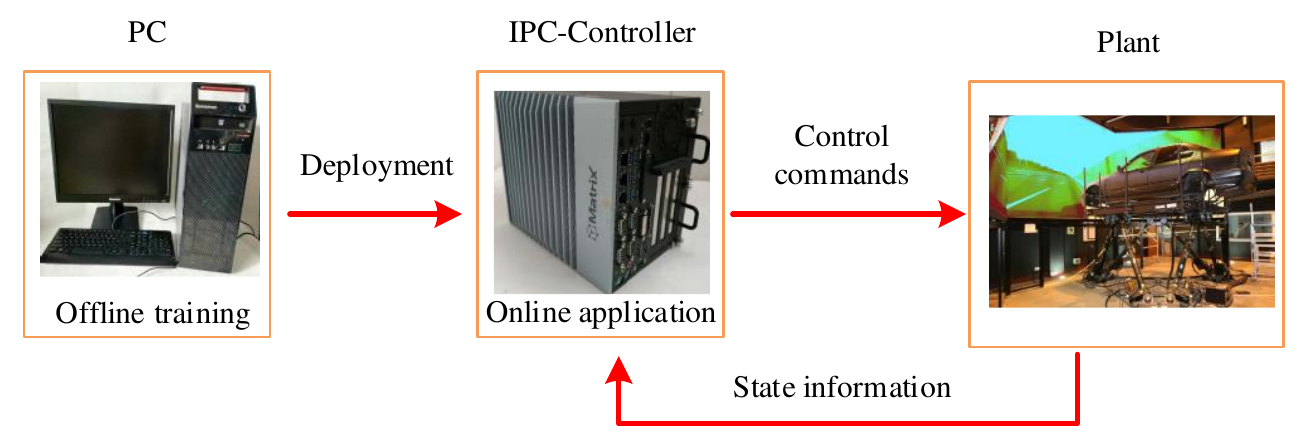}}
  \caption{Schematical view of the experimental setup.}
  \label{fig_HIL}
\end{figure}

The policy network is trained offline on the PC, and then deployed to the industrial personal computer (IPC). The vehicle dynamics used for policy training are different from the controlled plant. For online applications, the IPveC-controller gives the control signal to the plant according to the state information and the reference trajectory. The plant feeds back the state information to the IPC-controller, so as to realize the closed-loop control process. The feedback scheme of the HIL experiment is depicted in Fig \ref{fig_HIL}. The type of IPC-controller is ADLINK MXC-6401, equipped with Intel i7-6820EQ CPU and 8GB RAM, which is 
used as a vehicle on-board controller\cite{Chaoyi2019System}. The plant is a real-time system, simulated by the vehicle dynamic model of CarSim  \cite{carsim}.
The longitudinal speed is assumed to be constant, $v_x=16 \text{m/s}$, and the expected trajectory is shown in Fig. \ref{f:comparison_linear}. The system states and control inputs of this problem are listed in Table \ref{tab.state}, and the vehicle parameters are listed in Table \ref{tab.parameters}.

\begin{table}[!htb]
	\renewcommand{\arraystretch}{1.3}
	\caption{State and control input}
	\centering
	\label{tab.state}
	\resizebox{\columnwidth}{!}{
		\begin{tabular}{l l l l}
			\hline\hline \\[-3mm]
			\multicolumn{1}{c}{Mode} & \multicolumn{1}{c}{Name} & \multicolumn{1}{c}{Symbol}&
			\multicolumn{1}{c}{Unit}\\[1.6ex] \hline
		state&Lateral  velocity &$v_y$ & [m/s]  \\
		& Yaw rate at center of gravity (CG) &$\omega_r$ & [rad/s] \\
		& Longitudinal velocity &$v_x$ & [m/s] \\
		& Yaw angle &$\phi$ & [rad]\\
		& trajectory&$y$ & [m] \\
		input & Front wheel angle &$\delta$ & [rad] \\ 
			\hline\hline
		\end{tabular}
	}
\end{table}

\begin{table}[!htb]
	\renewcommand{\arraystretch}{1.3}
	\caption{Vehicle Parameters}
	\centering
	\label{tab.parameters}
	\resizebox{\columnwidth}{!}{
		\begin{tabular}{l l l}
			\hline\hline \\[-3mm]
			 \multicolumn{1}{c}{Name} & \multicolumn{1}{c}{Symbol}&
			\multicolumn{1}{c}{Unit}\\[1.6ex] \hline
Front wheel cornering stiffness &$k_1$ & -88000 [N/rad] \\
Rear wheel cornering stiffness &$k_2$ & -94000 [N/rad] \\
Mass &$m$ & 1500 [kg] \\
Distance from CG to front axle &$a$ & 1.14 [m] \\
Distance from CG to rear axle &$b$ & 1.40 [m] \\
Polar moment of inertia at CG &$I_z$ & 2420 [kg$\cdot\mathrm{m}^2$] \\
Tire-road friction coefficient &$\mu$ & 1.0 \\
Sampling frequency &$f$ & 20 [Hz] \\ 
System frequency & & 20 [Hz] \\ 
			\hline\hline
		\end{tabular}
	}
\end{table}


\subsection{Problem Description}
The offline policy is trained based on the nonliner and non input-affine vehicle dynamics:
\begin{equation}
\nonumber
x = 
\begin{bmatrix}
  y  \\
  \phi \\
  v_y \\
  \omega_r
\end{bmatrix}
,u =\delta 
,x_{i+1}=
\begin{bmatrix}
v_x \sin\phi + v_y \cos\phi\\
\omega_r\\
  \frac{F_{yf}\cos\delta + F_{yr}}{m} - v_x \omega_r\\
  \frac{aF_{yf}\cos\delta - bF_{yr}}{I_{z}}
   
\end{bmatrix}\frac{1}{f} + x_i,
\end{equation} 
where $F_{yf}$ and $F_{yr}$ are the lateral tire forces of the front and rear tires respectively \cite{kong2015kinematic}. The lateral tire forces can be approximated according to the Fiala tire model:
\begin{equation}
\nonumber
F_{y\#}  = 
\left\{
\begin{aligned}
&-C_\#\tan\alpha_\#\Big(\frac{C_\#^2(\tan\alpha_\#)^2}{27(\mu_\# F_{z\#})^2} -  \frac{{C_\#}\left |\tan\alpha_\# \right |}{3\mu_\# F_{z\#}} + 1\Big),\\ &\qquad \qquad \qquad \qquad \qquad \qquad |\alpha_\#|\le|\alpha_{\text{max},\#}|,\\
&  \mu_\# F_{z\#},&\\
&\qquad \qquad \qquad \qquad \qquad \qquad |\alpha_\#|>|\alpha_{\text{max},\#}|,
\end{aligned}
\right.
\end{equation} 
where $\alpha_{\#}$ is the tire slip angle, $F_{z\#}$ is the tire load, $\mu_\#$ is the friction coefficient, and the subscript $\# \in \{f,r\}$ represents the front or rear tires. The slip angles can be calculated from the relationship between the front/rear axle and the center of gravity (CG): 
\begin{equation}
\nonumber
\alpha_f = \arctan (\frac{v_y+a\omega_r}{v_x})-\delta, \quad \alpha_r = \arctan (\frac{v_y-b\omega_r}{v_x}). 
\end{equation} 
The loads on the front and rear tires can be approximated by: 
\begin{equation}
\nonumber
F_{zf} = \frac{b}{a+b}mg, \quad F_{zr} = \frac{a}{a+b}mg.
\end{equation} 

The utility function of this problem is set to be  $u_{i-1}$ 
\begin{equation}
\nonumber
l(x_i,{r_i},u_{i-1}) = ([1,0,0,0]x_i-r_i)^2+10{u_i}^2+([0,0,0,1]{x_i})^2.
\end{equation}

Therefore, the policy optimization problem of this example can be formulated as:
\begin{equation}
\nonumber
\centering
\begin{aligned}
\min_\theta  &  \mathop{\mathbb{E}}_{
\small\begin{array}{ccc}
\small x_0\in\mathcal{X}\\
\small r_{1:{N_{\text{max}}}}\in\mathcal{X}\\
\end{array} } \Big\{V(x_{0},r_{1:{N_{\text{max}}}},{N_{\text{max}}};\theta)\Big\}\\
s.t. \quad &x_{i} = f(x_{i-1},u_{i-1}), \\
&u_{\text{min}}\leq u_{i-1} \leq u_{\text{max}},\\
&i\in[1,N_{\text{max}}].
\end{aligned}
\end{equation}
where $V(x_{0},r_{1:{N_{\text{max}}}},{N_{\text{max}}};\textcolor{red}{\theta})=\sum_{i=1}^{N_{\text{max}}}l(x_i,r_i,u_{i-1})$, $u_{i-1}=\pi^{{N_{\text{max}}}-i+1}(x_{i-1},r_{i:{N_{\text{max}}}};\theta)$, $N_{\text{max}}=15$, $u_{\text{min}}=-0.2\text{rad}$ and $u_{\text{max}}=0.2\text{rad}$. 

\subsection{Algorithm Details}
The policy function is represented by a variant of RNN, called GRU (Gated Recurrent Unit). The input layer is composed of the states, followed by 4 hidden layers using rectified linear unit (RELUs) as activation functions with $128$ units per layer, and the output layer is set as a $tanh$ layer, multiplied by $0.2$ to confront bounded control. We use Adam method to update the network with the learning rate of $2\times10^{-4}$ and the batch size of $256$.

\subsection{Result Analysis}
For nonlinear MPC problems, we can solve it with some optimization solvers, such as ipopt \cite{Andreas2006Biegler} and bonmin \cite{bonami2008algorithmic}, which can be approximately regarded as the numerical optimal solution. 

\begin{figure}[!htb]
\captionsetup{justification =raggedright,
              singlelinecheck = false,labelsep=period, font=small}
\centering{\includegraphics[width=0.5\textwidth]{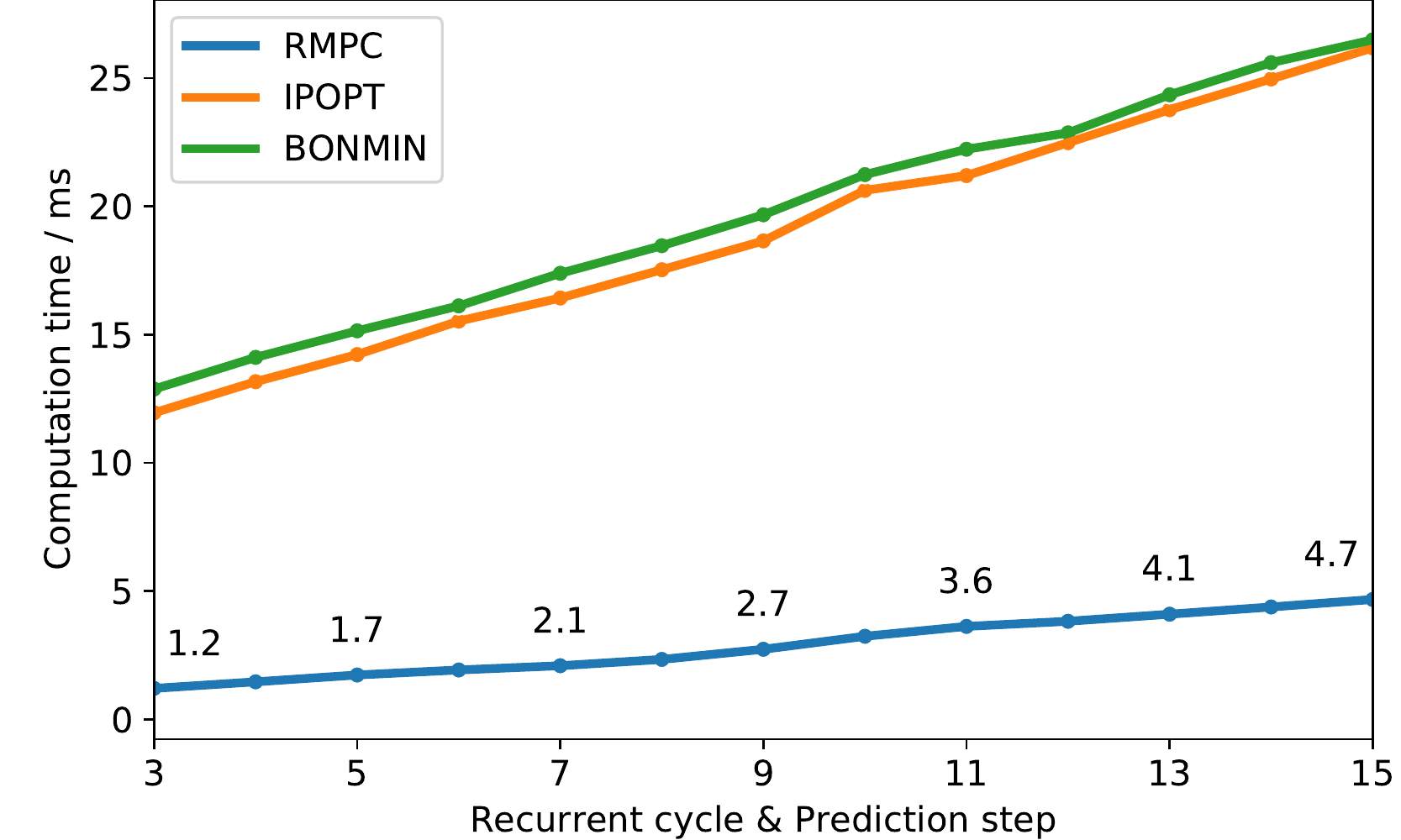}}
  \caption{RMPC vs optimization solvers computation time comparison.}
\label{fig_time}
\end{figure}

Fig. \ref{fig_time} compares the calculation efficiency of RMPC and the optimization solvers based on the symbolic framework CasADi \cite{2018CasADi} under different prediction steps for online applications. It is obvious that the calculation time of the optimization solvers is much longer than RMPC, and the gap increases with the number of prediction steps. Specifically, when $N=c=15$, the fastest optimization solver ipopt is over 5 times slower than RMPC (ipopt for $26.2$ms, RMPC for $4.7$ms). This demonstrates the effectiveness of the RMPC method.

We run Algorithm \ref{alg:RMPC} for 10 times and calculate the policy error $e_N$ between the solution of ipopt solver and RMPC at each iteration for  $N\in[1,15]$,
\begin{equation}
\nonumber
\begin{aligned}
e_N= \mathop{\mathbb{E}}_{
\small\begin{array}{ccc}
\small x_0\in\mathcal{X}\\
\small r_{1:N}\in\mathcal{X}\\
\end{array} }
&\left[
\frac{ \vert {u_{0}^{N}}^*(x_0,r_{1:N})-{\pi^N}(x_0,r_{1:N};\theta) \vert }{{u^N_{\text{max}}}^*-{u^N_{\text{min}}}^*}\right]\\
&\qquad \qquad,N\in[1,15],
\end{aligned}
\end{equation}
where ${{u^N_{\text{max}}}^*}$ and ${{u^N_{\text{min}}}^*}$ are respectively the maximum and minimum value of ${u_{0}^{N}}^*(x_0,r_{1:N})$ for $\forall x_0,r_{1:N} \in \mathcal{X}$, $\forall N \in [1,15]$, $N$ is the number of prediction steps. $e_N$ indicates the relative error of control quantity $\pi^N(\cdot)$  from cycle network respect to the optimum  $u_{0}^{N^*}(\cdot)$  in $N$ step prediction control problem .

\begin{figure}[!htb]
\captionsetup{
              singlelinecheck = false,labelsep=period, font=small}
\centering{\includegraphics[width=0.5\textwidth]{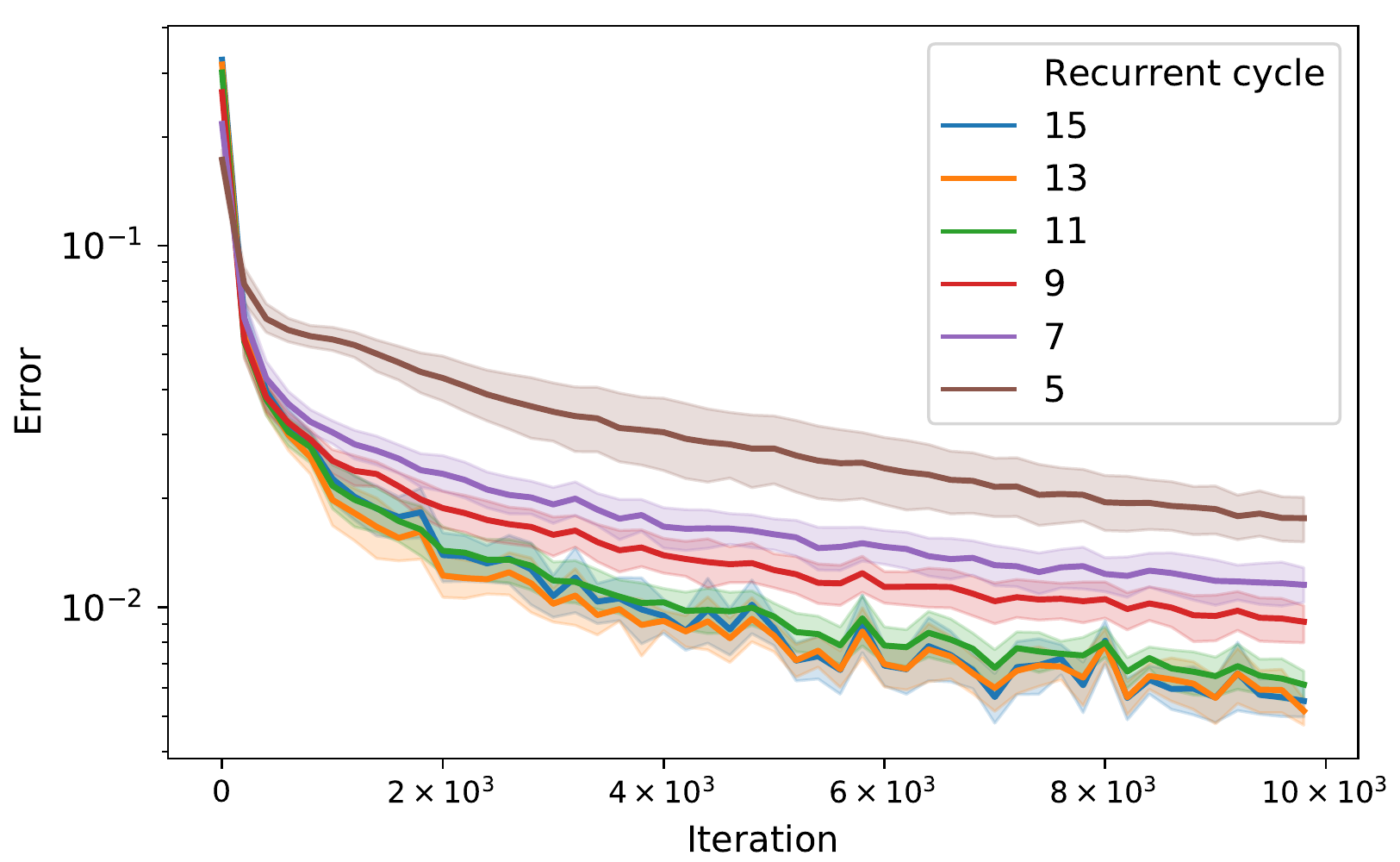}}
  \caption{Policy error during training. Solid lines are average values over 10 runs. Shaded regions correspond to 95\% confidence interval.}
  \label{fig_err_train}
\end{figure}
In Fig. \ref{fig_err_train}, we plot policy error curves during training with different prediction steps $N$ . It is clear that all the policy errors decrease rapidly to a small value during the training process. In particular, after $10^4$ iterations, policy errors for all  $N\geq5$ 
reduce to less than 2\%. This indicates that Algorithm  \ref{alg:RMPC} has the ability to find the near-optimal policy of MPC problems with different prediction horizons $N$.

\begin{figure}[!htb]
\captionsetup{
              singlelinecheck = false,labelsep=period, font=small}
\centering{\includegraphics[width=0.5\textwidth]{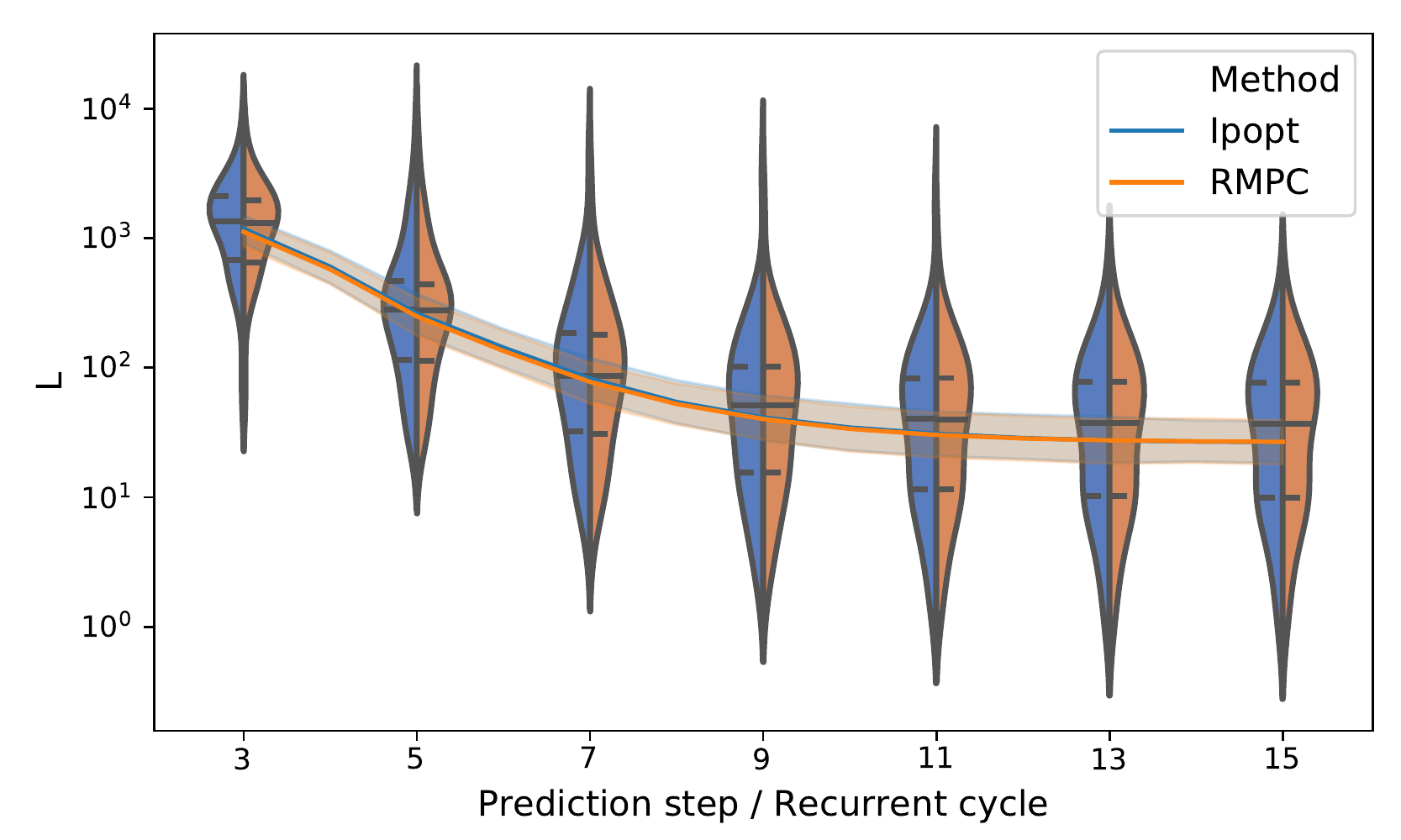}}
\caption{Performance comparison between the learned policy and the solution of ipopt solver. Solid lines are average values over 50 initialized states. Shaded regions correspond to 95\% confidence interval.}
\label{fig_loss}
\end{figure}

Fig. \ref{fig_loss} shows the policy performance of the ipopt solver solution and learned policy ${\pi^c}(x_0,r_{1:c};\theta)$ with different prediction horizons. The policy performance is measured by the lost function of 200 steps (10s) during the simulation period staring from random initialized state, i.e., 
\begin{equation}
\label{eq.loss_simulation}
\centering
L=\sum_{i=1}^{200} l(x_i,r_i,u_{i-1}). 
\end{equation}

For all prediction domains $N$, the learned policy performs as well as the solution of ipopt solver. More recurrent cycles (or long prediction steps) help reduce the accumulated cost $L$.

\begin{figure}[!htb]
\captionsetup{
              singlelinecheck = false,labelsep=period, font=small}
\centering{\includegraphics[width=0.5\textwidth]{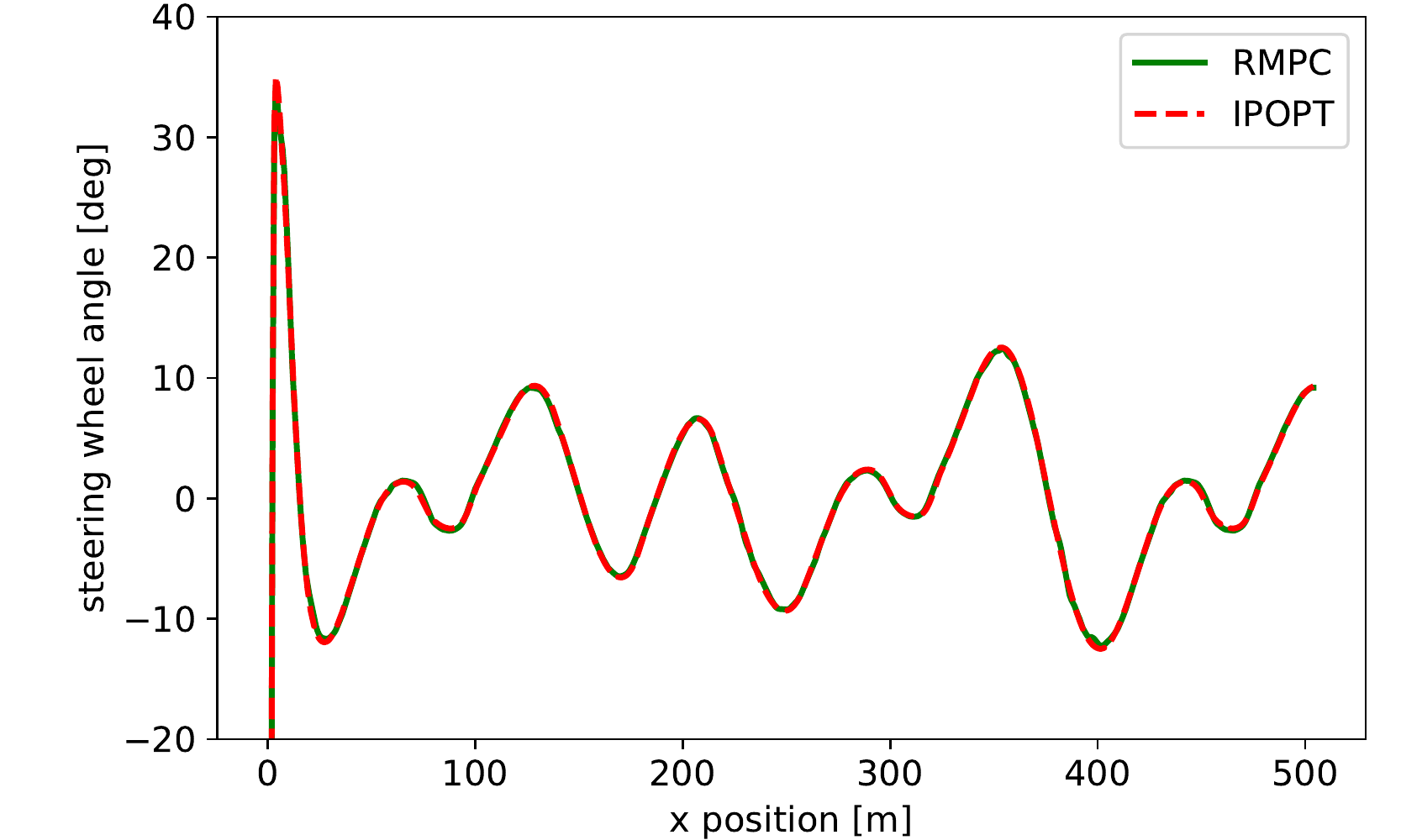}}
\caption{Control output comparison between the learned policy and the solution of ipopt solver, recurrent cycles $c=15$.}
\label{fig_output}
\end{figure}
\begin{figure*}[!htb]
\centering
\captionsetup{justification =raggedright,
              singlelinecheck = false,labelsep=period, font=small}
\captionsetup[subfigure]{justification=centering}
\subfloat[\label{subFig:l1}]{\includegraphics[width=0.31\textwidth]{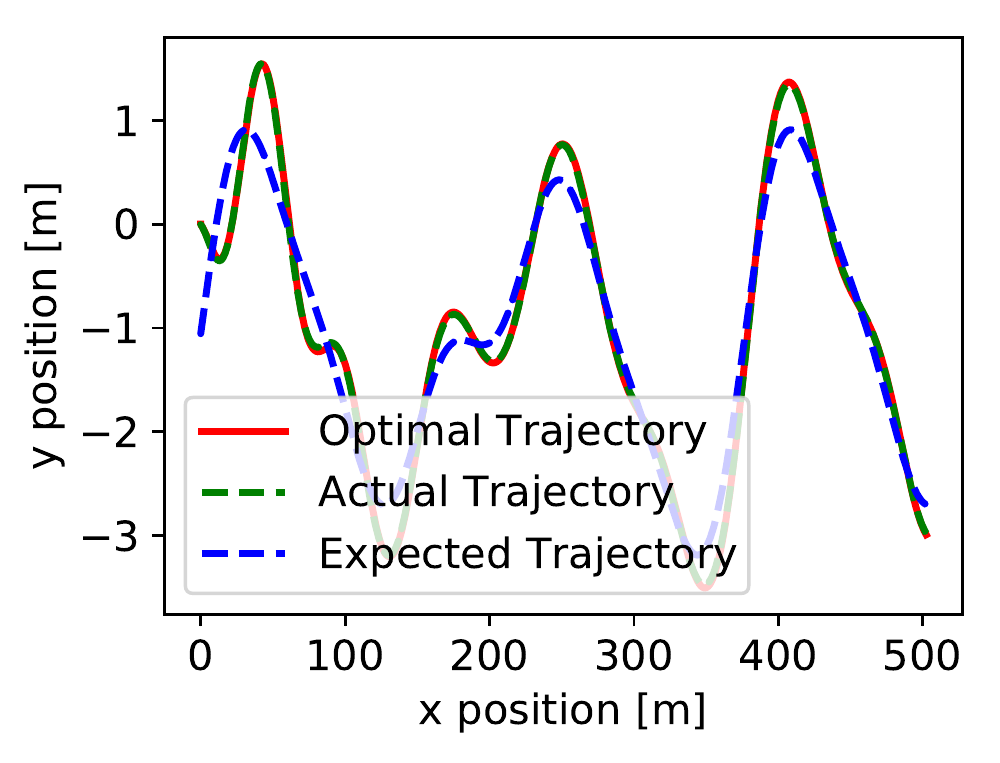}} 
\subfloat[ \label{subFig:r1}]{\includegraphics[width=0.31\textwidth]{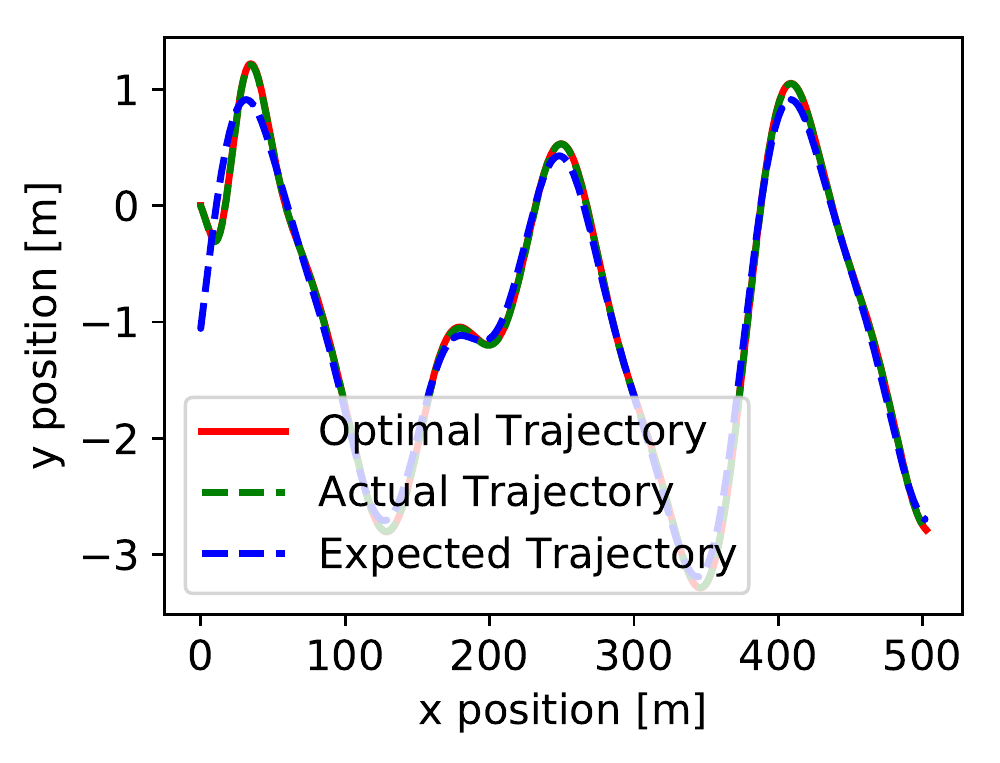}} 
\subfloat[\label{subFig:l2}]{\includegraphics[width=0.31\textwidth]{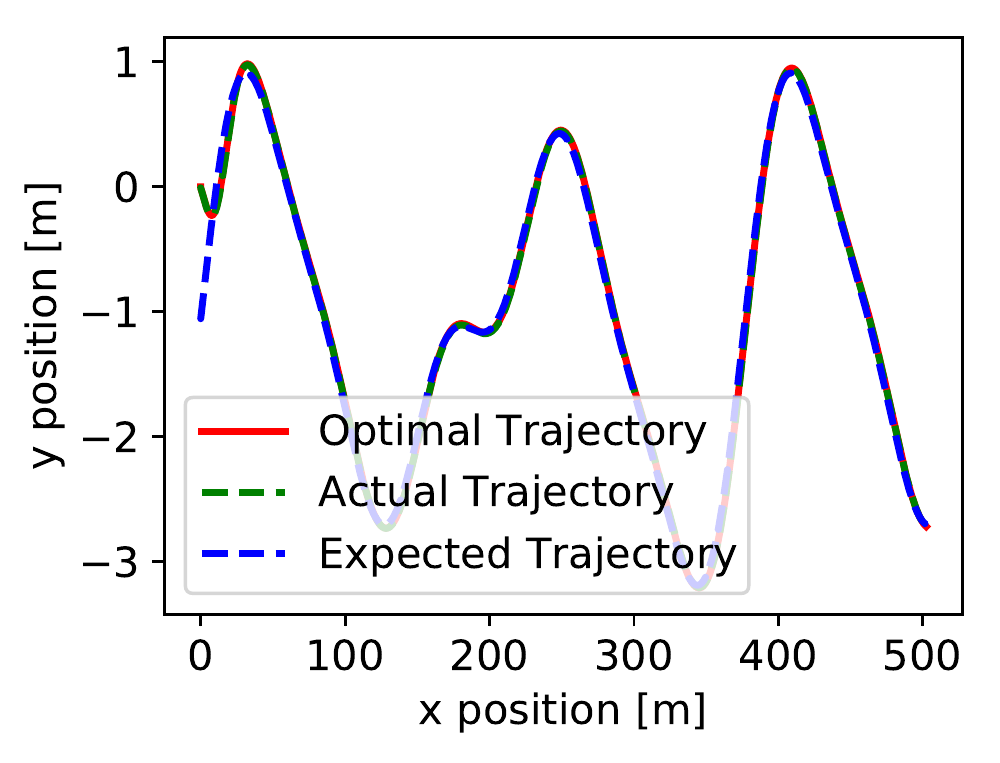}} \\
\subfloat[ \label{subFig:r2}]{\includegraphics[width=0.31\textwidth]{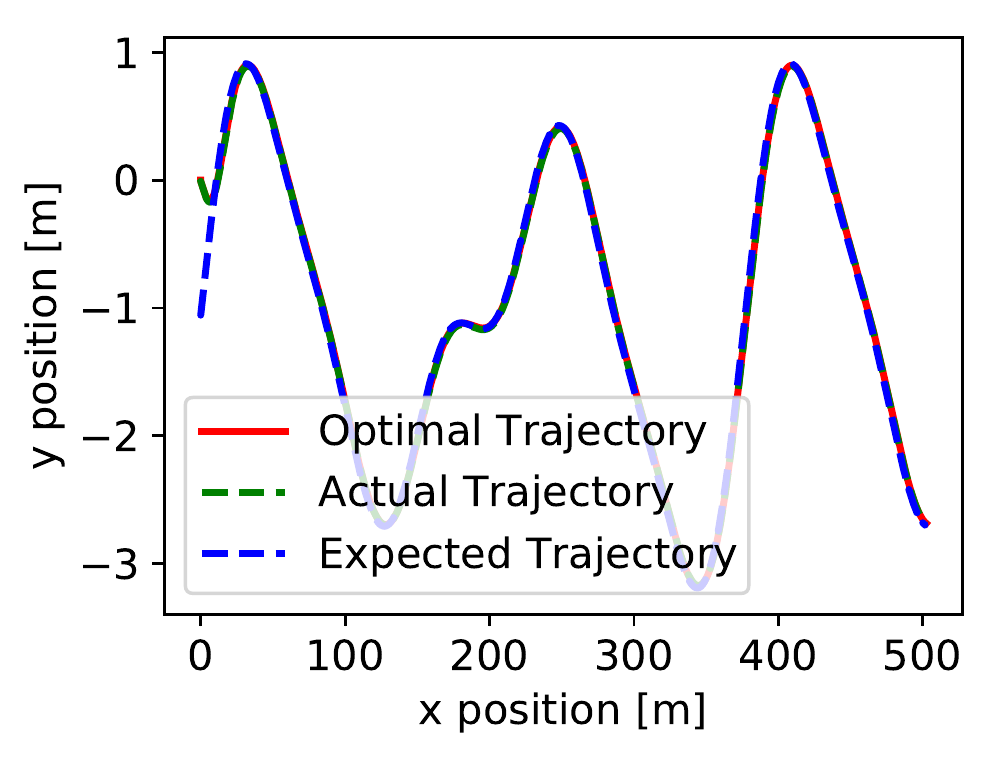}} 
\subfloat[\label{subFig:l3}]{\includegraphics[width=0.31\textwidth]{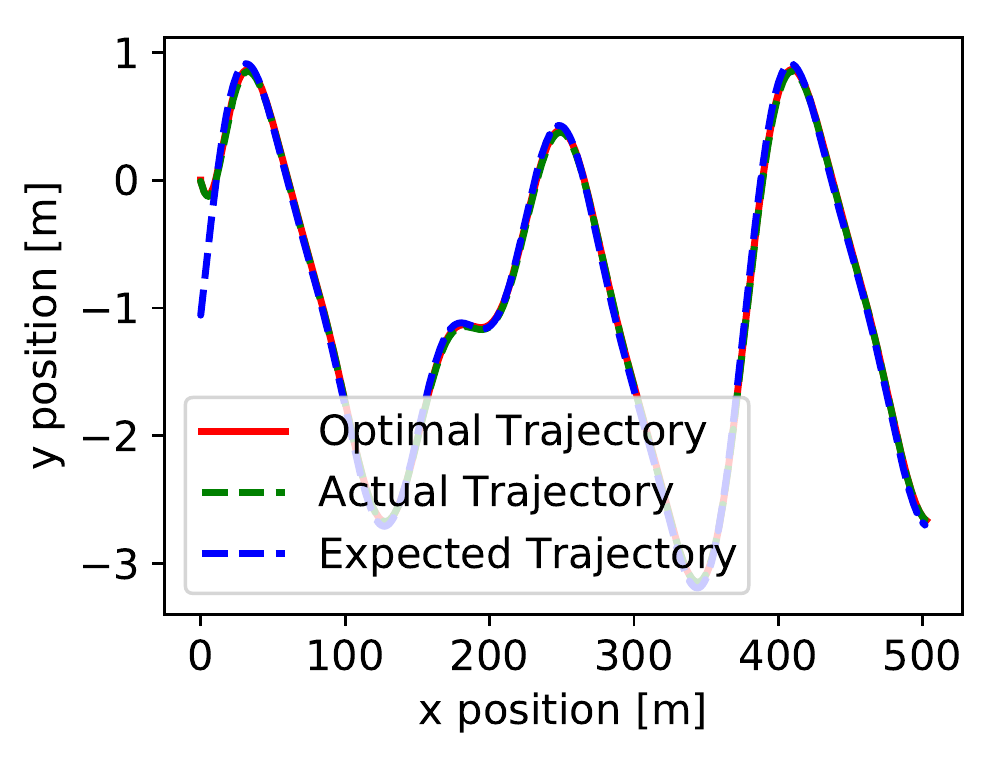}} 
\subfloat[\label{subFig:r3}]{\includegraphics[width=0.31\textwidth]{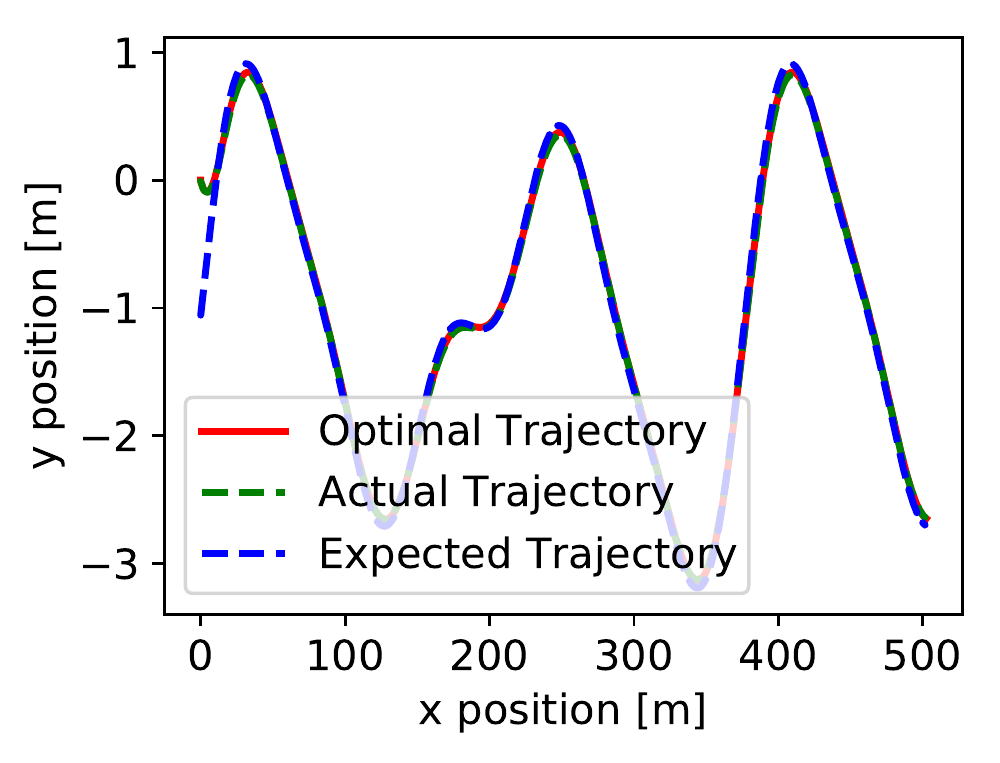}} 
\caption{Tracking results for policies with different recurrent cycles $c$. (a) $c=5$. (b) $c=7$. (c) $c=9$. (d) $c=11$. (e) $c=13$. (f) $c=15$.}
\label{f:comparison_linear}
\end{figure*}

In detail, Fig. \ref{f:comparison_linear} intuitively presents the control results of the learned policy with different recurrent cycles $c$ and  Fig. \ref{fig_output} compares the control output between the learned policy(after 15 recurrent cycles) with ipopt controller.Obviously, the trajectory controlled by RMPC controller almost overlaps with the ipopt controller. The more recurrent cycles of the learned policy, the smaller the trajectory tracking error.This is why we want to adaptively select the optimal law with longest prediction horizon in real applications.

To summarize, the example demonstrates the optimality, efficiency and generality of the RMPC algorithm.

\section{Conclusion}
This paper proposes the Recurrent Model Predictive Control (RMPC) algorithm to solve general nonlinear finite-horizon optimal control problems.
Unlike traditional MPC algorithms, it can make full use of the current computing resources and adaptively select the longest model prediction horizon. Our algorithm employs an RNN to approximate the optimal policy, which maps the system states and reference values directly to the control inputs. The output of the learned policy network after $N$ recurrent cycles corresponds to the nearly optimal solution of $N$-step MPC. A policy optimization objective is designed by decomposing the MPC cost function according to the Bellman's principle of optimality.The optimal recurrent policy can be obtained by directly minimizing the designed objective function, which is applicable for general nonlinear and non input-affine systems. The convergence and optimality of RMPC is further proved. We demonstrate its optimality, generality and efficiency using a HIL experiment. Results show that RMPC is over 5 times faster than the traditional MPC algorithm. The control performance of the learned policy can be further improved as the number of recurrent cycles increases.

\newpage
\bibliographystyle{./Bibliography/IEEEtranTIE}
\bibliography{./Bibliography/IEEEabrv,Bibliography/BIB_xx-TIE-xxxx}

\end{document}